\documentclass[]{spie}  

\usepackage{amsmath,amsfonts,amssymb}
\usepackage{graphicx}
\usepackage[colorlinks=true, allcolors=blue]{hyperref}

\title{Ferromagnetic Josephson junctions for cryogenic memory}

\author[a]{Norman O. Birge}
\author[a]{Alexander E. Madden}
\affil[a]{Michigan State University, Dept. of Physics and Astronomy, East Lansing, MI 48823, USA}
\author[b]{Ofer Naaman}
\affil[b]{Northrop Grumman Systems Corp., Baltimore, MD 21240, USA}
\authorinfo{N.O.B. e-mail: birge@pa.msu.edu\\O.N. current address: Google Inc., Santa Barbara, California 93117, USA}

\begin{document}
\maketitle

\begin{abstract}
Josephson junctions containing ferromagnetic materials have attracted intense interest both because of their unusual physical properties and because they have potential application for cryogenic memory.  There are two ways to store information in such a junction: either in the amplitude of the critical current or in the ground-state phase difference across the junction; the latter is the topic of this paper.  We have recently demonstrated two different ways to achieve phase control in such junctions: the first uses junctions containing two magnetic layers in a ``pseudo spin valve" configuration, while the second uses junctions containing three magnetic layers with non-collinear magnetizations.  The demonstration devices, however, have not yet been optimized for use in a large-scale cryogenic memory array.  In this paper we outline some of the issues that must be considered to perform such an optimization, and we provide a speculative ``phase-diagram" for the nickel-permalloy spin-valve system showing which combinations of ferromagnetic layer thicknesses should produce useful devices.
\end{abstract}


\section{INTRODUCTION}
\label{sec:intro}  

Josephson junctions containing ferromagnetic materials have been under intense study by several groups since the pioneering work of the Ryazanov and Aprili groups nearly 20 years ago.\cite{Ryazanov2001,Kontos2002}  Those two papers were the first to report observation of ferromagnetic ``$\pi$-junctions'', which had first been predicted 20 years earlier.\cite{Buzdin1982,Buzdin1991}  A conventional Josephson junction has a current-phase relation of the form $I_s = I_c$sin$(\phi)$, where $I_s$ is the supercurrent, $I_c$ is the critical current (the maximum current that can flow through the junction with no voltage drop), and $\phi$ is the difference between the phases of the superconducting condensates in the electrodes on either side of the junction.  Such a junction has a minimum energy when $\phi = 0$.  In contrast, $\pi$-junctions have an inverted current phase relation, $I_s = I_c$sin$(\phi+\pi) = -I_c$sin$(\phi)$, and minimum energy when $\phi = \pi$.  Such $\pi$-junctions are interesting in their own right, and have also attracted interest due to their potential use as circuit elements in both ``single-flux-quantum" superconducting circuits and in superconducting quantum computing circuits.

In this paper we focus on one particular circuit application, namely cryogenic memory.  Proposals for cryogenic memory date back at least to the 1970's when IBM was trying to develop a superconducting computer.\cite{Faris1980}  Progress on cryogenic memory continued in Japan after the IBM program was canceled,\cite{Miyahara1984} then the field received renewed energy with the development of ``single-flux-quantum" (SFQ) logic at Moscow State University in the late 1980's.\cite{LikharevSemenov1991}   By the mid-1990s the Japanese had demonstrated a fully-functional 4 kbit memory array with integrated drivers to enable memory access by an SFQ processor.\cite{Nagasawa1995}  Further progress in SFQ logic and memory has waxed and waned over the years according to the availability of government funding.  The early hopes that superconducting computers would beat semiconductor-based computers in speed and performance were largely dashed by the spectacular advances in the performance of CMOS-based systems over the past three decades.  Recently, however, there is renewed interest in superconducting computing due to the immense amount of electrical power being used nowadays by large-scale computing systems and data centers.  That interest has been fueled by the development of several new forms of SFQ logic that are much more energy-efficient than the original formulation.\cite{Kirichenko2011,Volkmann2013,Herr2011,Tanaka2012,Takeuchi2013}  Given these recent developments, some observers believe that a superconducting computer, while not necessarily out-performing semiconductor-based computers in terms of speed, could function at comparable speeds while consuming much less power.\cite{Holmes2013,Soloviev2017_review}

Success in developing superconducting computing requires the development of new forms of cryogenic memory.  The standard way to make memory compatible with SFQ circuits involves storing a quantum of magnetic flux, $\Phi_0 = h/2e = 2.0 \times 10^{-15}$T-m$^2$, with a circulating current in a superconducting loop.  Since typical values of Josephson junction critical currents in SFQ circuits are of order $I_c \approx 100 \mu$A, the inductance of the storage loop must be of the order of 20 pH.\cite{Mannhart2006}  To achieve such an inductance with a planar loop requires the loop diameter to be larger than about 10 $\mu$m.  In other words, a memory based solely on SFQ technology is not scalable to high density.

There have been several proposals to use ferromagnetic materials as the memory-containing element in a superconducting memory.  These can be roughly separated into two categories: 1) those that use the magnetic field produced by a nanomagnet to modulate the critical current of a nearby superconducting device such as a nanowire, Josephson junction, or SQUID;\cite{Mannhart2006,Johnson1999} and 2) those that use the exchange field of the nanomagnet either to modulate the properties of a nearby superconducting element via the proximity effect,\cite{Beasley1997} or to modulate the properties of a Josephson junction containing the nanomagnet.\cite{Bell2004}  A generic feature of the first category of proposals is that the magnitude of the effect gets weaker as the device size shrinks.  To illustrate that, consider proposals to modulate the properties of a Josephson junction with a nearby magnetic field, via the ``Fraunhofer effect."  Such modulation requires introducing a quantum of magnetic flux, $\Phi_0 = h/2e$, into the junction.  The relevant area over which the field induces flux is typically equal to twice the London penetration depth times the junction dimension in the direction perpendicular to the magnetic field.  As that dimension shrinks, the required field magnitude grows.  Schemes that use the exchange field do not suffer from that drawback.  Those can be further divided into proposals to control the critical temperature of a superconducting wire, or proposals to control the supercurrent and/or phase properties of a Josephson junction.  The former suffer from the drawback that the sample temperature must be maintained in a very narrow range close to the critical temperature of the device.\cite{Leksin2010}  In contrast, memory schemes involving Josephson junctions usually work at a convenient fixed temperature such as 4.2 K.  In this paper we will consider only this last sub-category of devices.

\section{Physics of ferromagnetic Josephson junctions }

The physics of Josephson junctions containing ferromagnetic materials has been reviewed several times.\cite{Lyuksyutov2005,Buzdin2005,Bergeret2005}  The basic idea can be summarized as follows: a Cooper pair in one of the superconducting (S) electrodes enters the ferromagnetic layer (F).  Assuming that S is a conventional superconductor with spin-singlet pairs, the two electrons cannot enter the same spin band in F.  Instead, one electron enters the majority band with momentum $+\hbar k_F^{\uparrow}$, while the other enters the minority band with momentum $-\hbar k_F^{\downarrow}$.  The pair thereby gains a momentum, $\hbar Q = \hbar (k_F^{\uparrow}-k_F^{\downarrow})$, or equivalently, the pair correlation function oscillates in space with wavevector $Q$.\cite{Demler1997}  If the F-layer thickness is such that the pair correlation function is negative when it reaches the second S electrode, then the Josephson coupling between the two S electrodes will be negative, leading to a $\pi$-junction.  If a series of samples is fabricated with a range of F-layer thicknesses $d_F$, the critical current, $I_c$, is observed to oscillate in magnitude as a function of $d_F$.  Direct measurements of the current-phase relation verify that junctions with $d_F$ in the appropriate range are in fact $\pi$-junctions.\cite{Frolov2004}

While Josephson $\pi$-junctions with a single F layer may be useful in their own right, the situation gets much more interesting if the properties of a single junction can be changed in-situ.  This can be achieved if the junction contains more than one F layer.\cite{Bergeret2001PRL,Krivoruchko2001,Golubov2002}  In particular, it was predicted by Golubov \textit{et al.} in 2002 that a junction containing two magnetic layers could be switched between the 0 and $\pi$ states by changing the mutual orientation of the F layers from parallel to antiparallel.\cite{Golubov2002}  A couple of years later, Bell \textit{et al.} suggested that such a ``spin-valve" junction could be used as a memory device.\cite{Bell2004}  Workers at Northrop Grumman Corporation (NGC) seized upon this idea and developed a memory architecture that enables read and write addressing for a large-scale memory array.\cite{herr_patent2012}  The NGC scientists call their architecture JMRAM, for ``Josephson magnetic random access memory."  Our own work on cryogenic memory has largely focused on the development of JMRAM.  We note that, in principle, either the amplitude of $I_c$ or the ground-state phase of the junction can be used as the fundamental memory storage mode.  In practice, however, there are advantages to using the phase.\cite{herr_phasepatent2015}  If the amplitude of $I_c$ is the quantity of interest, then the readout mechanism will necessarily require switching the ferromagnetic junction into the voltage state when it is in the state with lower $I_c$.  But Josephson junctions containing multiple F layers tend to have small values of $I_c R_N$ -- the product of critical current times normal-state resistance.  Hence the voltage signal generated by the junction switching into the voltage state will be small, and the switching time will be long, $t_{switch} \approx \hbar/(2eI_c R_N)$.  Instead, one can use the phase of the junction for memory storage by placing the ferromagnetic junction in a SQUID loop, along with two conventional superconductor-insulator-superconductor (SIS) junctions that have critical currents smaller than that of the ferromagnetic junction.\cite{Dayton2018}  When the ferromagnetic junction is in the zero state, the critical current of the SQUID is equal to the sum of the two SIS junction critical currents.  When the ferromagnetic junction is in the $\pi$-state, the critical current of the SQUID is equal to the difference between the two SIS critical currents.  In this scheme the ferromagnetic junction acts as a passive phase shifter, and always remains in the zero-voltage state.  Readout of the SQUID critical current involves switching of one of the SIS junctions, which have large values of the $I_c R_N$ product.

Before discussing the details of the spin-valve Josephson junctions, we mention that there is another type of ferromagnetic Josephson junction that allows phase control.  Junctions containing three F layers with non-collinear magnetizations between adjacent F layers can carry spin-triplet supercurrent.  Such junctions have generated widespread interest partly because spin-triplet supercurrents can be spin polarized, leading to the tantalizing possibility of a new field of "superconducting spintronics."\cite{LinderRobinson2015,Eschrig2015}  We recently demonstrated that spin-triplet junctions can exhibit in-situ phase control, by changing the orientation of one of the three F layers while keeping the other two fixed.\cite{Glick2018}  Many of the considerations regarding optimization of spin-valve junctions for cryogenic memory also apply to spin-triplet junctions.  Due to the presence of a third layer, however, the latter have additional constraints.  In this paper we will focus on spin-valve junctions.

\section{Optimization of spin-valve junctions for cryogenic memory}

We reported the demonstration of a controllable 0-$\pi$ spin-valve Josephson junction in 2016,\cite{Gingrich2016} and demonstration of a JMRAM unit cell containing such a junction was reported by NGC earlier this year.\cite{Dayton2018}  Although both of those works used junctions containing a Ni ``fixed layer" and a permalloy (NiFe) ``free layer," the physical specifications of the junctions used in the two works were quite different.  In our work,\cite{Gingrich2016} the Ni and NiFe thicknesses were 1.2 nm and 1.0 nm, respectively, whereas in the NGC work,\cite{Dayton2018} the two thicknesses were 3.0 nm and 1.6 nm, respectively.  In addition, our junctions had lateral dimensions of approximately $0.5 \mu$m$\times 1.25 \mu$m, whereas the NGC junctions had lateral dimensions of about $1.0 \mu$m$\times 2.0 \mu$m.  Several questions naturally arise regarding how these choices were made and whether they are optimal in any sense.

Optimizing the performance of JMRAM involves two completely separate sets of specifications for the magnetic and superconducting properties, respectively.  Regarding the magnetic properties, we want the fixed magnetic layer in the junction to have a high switching field, and preferably to be monodomain to minimize the appearance of stray field at domain walls, which might have deleterious effects on the behavior of the free layer.  In contrast, we want the free magnetic layer to have as low a switching field as possible, with very little sample-to-sample variation, while also being monodomain.  Regarding the superconducting properties, the main consideration is that the critical current of the Josephson junction be large enough so that the junction always stays in the supercurrent state during the read operation acting on the cell.  In practice, that means that the critical current should be significantly larger than about 100 $\mu$A, which is a typical critical current for the SIS junctions in the memory cell SQUID loop.  There are several design choices that affect either the magnetic or superconducting properties of ferromagnetic Josephson junctions, or both.  In this paper we discuss four: 1) choice of ferromagnetic materials and nonmagnetic spacers; 2) smoothness of the underlying superconducting electrode; 3) lateral dimensions of the junctions; and 4) thicknesses of the ferromagnetic materials.  One choice that has remained fixed in current research is the choice of superconducting material.  Essentially all of the modern work on cryogenic memory uses Nb as the superconductor, for several reasons.  Bulk Nb has a critical temperature of 9.1 K, allowing operation of Nb-based circuits at the liquid helium boiling temperature of 4.2 K.  Josephson junctions made from the standard Nb-Al-Al$_2$O$_3$-Nb trilayer process have excellent and highly reproducible characteristics.  Nb-based superconducting circuits, including the Josephson junctions, are also extremely robust to thermal cycling.  For those and other reasons, Nb is the material of choice of nearly every group working in the area of superconducting electronics.

One disadvantage of working with Nb is that sputtered Nb films grow with a ``rice-grain texture" that gets rougher with increasing film thickness.  Growing thin magnetic films on top of a rough Nb base layer generally leads to poor magnetic properties.  Our group has partially alleviated that problem by following an old recipe that uses an Al/Nb multilayer in place of the pure Nb bottom electrode.\cite{Wang2012,Thomas1998}  We have found that magnetic films grown on the multilayer have better magnetic properties than similar films grown on thick Nb. Other groups have reached similar conclusions using other methods of planarization, such as oxidation and ion milling\cite{Singh2018} or deposition of an Al/Cu seed bilayer between the Nb and the ferromagnetic layer.\cite{Loving2018}  This issue is an area of ongoing research.

The number of different ferromagnetic materials that have been studied inside Josephson junctions so far is only about a dozen, but the number continues to grow.  These include ``weak" ferromagnetic alloys such as CuNi,\cite{Ryazanov2001} PdNi,\cite{Kontos2002} or PdFe,\cite{Ryazanov2012,Glick2017PdFe} as well as ``strong" ferromagnetic elements and alloys such as Ni,\cite{Blum2002,Shelukhin2006} Fe,\cite{Robinson2006} Co,\cite{Robinson2006} NiFe,\cite{Robinson2006,Glick2017} NiFeNb,\cite{Baek2014} NiFeMo,\cite{Niedzielski2015} and NiFeCo.\cite{Glick2017}  These materials differ in a number of respects, both with respect to their magnetic properties and their superconducting properties.  Furthermore, the magnetic properties of most of these materials are not strictly intrinsic, but depend also on the environment in which they are placed (e.g. substrate smoothness) and on their thickness.  For example, permalloy (Ni$_{80}$Fe$_{20}$) is a well-known and often-used soft magnetic material, with extremely low magnetocrystalline anisotropy and magnetostriction in the bulk at room temperature.  The coercive field $H_c$ of a permalloy film several tens of nm thick can be or order 0.1 Oe.  A permalloy film only 1 nm thick, in contrast, may have a coercive field of 1 Oe at room temperature, increasing to a few Oe at low temperature.  Furthermore, if a similar film is grown on a rough underlayer, $H_c$ increases significantly.  We note that the mechanism of magnetization reversal in a continuous thin film is very different from the reversal process in a small patterned magnetic ``bit", if the bit is single-domain.  In the former, the reversal typically occurs by domain wall motion, whereas in the latter the reversal occurs by coherence rotation of the magnetization, also called ``Stoner-Wohlfarth" switching.  Hence the value of $H_c$ in a film is typically determined by the extent of domain wall pinning by defects, while the switching field of a nanomagnet in the ideal case is determined by the shape anisotropy of the bit.  In practice, however, the nanomagnet switching field will vary from bit to bit due to uncontrolled extrinsic factors such as defects or roughness.  For example, we have found that the switching distribution for a large array of permalloy bits moves to higher average field and broadens when the permalloy is grown on thick Nb (100 nm) compared to when it is grown on a very thin Nb (5 nm).\cite{Bethanythesis}  Since the extrinsic factors that broaden the switching distribution of bit arrays also tend to increase $H_c$ and broaden the switching transition of continuous films (i.e. reduce the ``squareness" of the transition), efforts to minimize $H_c$ in continuous films should help in reducing the width of the switching distribution in nanomagnet arrays.

Returning to the list of ferromagnetic materials studied in Josephson junctions, let us consider the advantages and disadvantages of using ``weak" vs ``strong" ferromagnetic materials.  There are two obvious advantages of using weak ferromagnetic materials.  From the magnetic side, Stoner-Wohlfarth theory predicts that the switching field of a single-domain nanomagnet is proportional to its magnetization, and the switching energy is proportional to the square of the magnetization.  Hence lower magnetization should mean lower switching field.  In practice, however, one achieves the Stoner-Wohlfarth value of the switching field in a nanomagnet only if extrinsic sources of anisotropy are minimized.  From the superconducting side, the characteristic length scale $\xi_F$ over which junctions oscillate between being 0-junctions and $\pi$-junctions as the F-layer thickness changes is very short in strong ferromagnetic materials.  In the ballistic transport regime, $\xi_F = Q^{-1} \approx \hbar v_F/2E_{ex}$, where $Q = k_F^{\uparrow}-k_F^{\downarrow}$ was discussed earlier, and the approximate equality comes from an oversimplified parabolic band model of the F material with Fermi velocity $v_F$.  In the diffusive transport regime, $\xi_F = (\hbar D/E_{ex})^{1/2}$, where $D$ is the diffusion constant.  In strong ferromagnetic materials such as Fe or Co, $E_{ex}$ is of order 1 eV and $\xi_F$ is typically less than 1 nm.\cite{Robinson2006}  With such a short length scale governing the junction properties, those properties may fluctuate if the average F-layer thickness varies by even a fraction of an atomic monolayer.  Weak F materials typically have much larger values of $\xi_F$, so it is easier to control the sample thicknesses and sample-to-sample variations in junction properties are generally smaller.\cite{Glick2017,Oboznov2006}  Given these two strong advantages, why haven't weak F materials dominated the landscape of research into cryogenic memory?  One reason is that some weak F materials severely depress the critical current when they are inserted into Josephson junctions.  This issue was apparent already in the early work of the Ryazanov group on the Cu$_{1-x}$Ni$_x$ alloy, where they found a steep decrease in $I_c$ as a function of increasing CuNi thickness.\cite{Oboznov2006}  More recently, Baek \textit{et al.}\cite{Baek2014} have shown that a Nb-doped permalloy with low magnetization has excellent magnetic switching properties at low temperature.  We also looked into doping permalloy with molybdenum to lower its magnetization.  Unfortunately, we have found that both Nb-doped and Mo-doped permalloy severely suppress the critical current in Josephson junctions.\cite{Niedzielski2014,Niedzielski2015}  A striking exception to this trend is seen in the Pd-based alloys, Pd$_{1-x}$Ni$_x$ and Pd$_{1-x}$Fe$_x$.  Josephson junctions containing these alloys can have very high critical currents.  The former is less attractive for us because the of a high coercive field and because the magnetization has a tendency to align out-of-plane.  But PdFe avoids both of those drawbacks.  Pd$_{1-x}$Fe$_x$ alloys are ferromagnetic even with extremely small Fe concentrations.  Ryazanov and collaborators at Hypres have performed numerous studies of Pd$_{1-x}$Fe$_x$ with $x = 1\%$.\cite{Ryazanov2012,Mukhanov2012}  That alloy has an extremely low Curie temperature of about 10 K, and low coercive field.  Detailed analysis of the Fraunhofer pattern of a 4-$\mu$m square junction shows, however, that the magnetization switching takes place over a broad field range starting near 0 field and ending at about 30 Oe. Our own work on small junctions containing PdFe with 3\% Fe also shows that the magnetization switching is not abrupt, but rather occurs over a range of fields ending somewhere between 20 and 30 Oe.\cite{Glick2017PdFe}  Of more concern is that the hysteresis curve is reproducible only if the applied field exceeds about 50 Oe.  It is still too early to know if PdFe will be a useful material for the free layer in a spin-valve Josephson junction.

After the preceding discussion, it may come as a surprise that the best experimental results on junction control have come from Josephson junctions containing Ni as the fixed layer and permalloy as the free layer -- both of those strong ferromagnetic materials.  The choice of permalloy is somewhat obvious given its excellent switching properties at low field.  The choice of Ni is largely due to its excellent supercurrent-carrying capacity.  S/F/S Josephson junctions containing Ni exhibit values of $I_c R_N$ in the $\pi$ state of order 100 $\mu$V or possibly larger.\cite{Baek2017}  That should be compared with S/F/S junctions containing permalloy or Cu$_{47}$Ni$_{53}$ alloy, which exhibit maximum $I_c R_N$ values in the $\pi$ state of about 15 $\mu$V or 1 $\mu$V, respectively.\cite{Glick2017,Oboznov2006}  Furthermore, theoretical fits to $I_c$ vs $d_{Ni}$ data indicate that Ni is in the ballistic (clean) limit rather than the diffusive (dirty) limit, hence $I_c$ decays only algebraically with increasing $d_{Ni}$, rather than exponentially.\cite{Baek2017}  The drawback of using Ni is that its magnetic properties are not ideal.  The good news is that the Ni coercivity is high, but the bad news is that the Ni nanomagnets in the current generation of devices are probably not monodomain.  Switching of a 2.0-nm thick Ni layer in an elliptical Josephson junction with lateral dimensions $0.5 \mu\textrm{m}\times 1.25 \mu\textrm{m}$ takes place over an extended field range of about 400-1000 Oe.\cite{Niedzielski2018}  The best results on S/F/S junctions containing only Ni were obtained after initializing the Ni with a field of 3500 or 4000 Oe.\cite{Baek2017}  As long as subsequent applied field are kept below 100 Oe or so, the Ni magnetization stays nearly constant.  The successful demonstrations of controllable 0-$\pi$ switching and JMRAM unit cell performance have been carried out using a similarly high initialization field, with subsequent switching fields never exceeding 100 Oe.

The above discussion immediately brings up another issue, namely junction size.  Surely if the junctions were made small enough, the Ni nanomagnets would be single-domain, and might have more reliable magnetic properties.  The main drawback to reducing junction size is that $I_c$ scales with size.  As discussed earlier, in the current generation of devices $I_c$ of the ferromagnetic junction is determined by the $I_c$ values of the SIS junctions, which are in the range of 100$\mu$V.  Unless we find a way to increase the critical current density, $J_c = I_c/area$, we cannot decrease the lateral dimensions much.  A second issue with decreasing junction size is that the magnetostatic interaction between the fixed and free magnetic layers increases in relative importance as size decreases.  The magnetostatic interaction favors the antiparallel magnetic state over the parallel state.  That issue could be overcome by applying a constant ``bias" field in the direction of the fixed layer magnetization, but one would prefer to avoid the presence of large-scale magnetic fields near superconducting circuits.

\section{Tentative ``phase diagram" for Ni/NiFe spin-valve Josephson junctions}

At last we come to the main topic of this paper, namely how to choose the thicknesses of the ferromagnetic layers to ensure that the parallel and antiparallel magnetic states correspond to different phase states of the junction, i.e. either the $\pi$ and 0 phase states or vice versa.  This is the issue where it would seem that theoretical calculations would be most beneficial.  The problem is that the actual devices used by the experimental groups are much more complicated than even the most sophisticated theoretical models put forward to date.  One feature of the current generation of devices that challenges theoretical modeling is the mix of Ni and permalloy.  The behavior of Ni in a Josephson junction by itself is described well by a theoretical formula valid in the ballistic limit.\cite{Baek2017}  The behavior of permalloy in a junction by itself is described by a theoretical formula valid in the diffusive limit.\cite{Glick2017}  Baek \textit{et al.}\cite{Baek2017} have shown that, when a strongly-scattering layer is added to a Josephson junction containing Ni, the position of the 0-$\pi$ transition shifts to a larger Ni thickness, even if that additional scattering layer is non-magnetic.  While that shift is not fully understood, it may arise primarily due to a change in the transport properties of the junction from predominantly ballistic to predominantly diffusive.  This hypothesis is based on the following argument: in a ballistic SFS Josephson junction with no potential barriers at the interfaces, the first 0-$\pi$ transition occurs at $d_F/\xi_F \approx \pi/4$.\cite{Buzdin1982}  In a diffusive junction with no potential barriers, the first the first 0-$\pi$ transition occurs at $d_F/\xi_F \approx 3\pi/4$, where now we are referring to the diffusive form of $\xi_F$.\cite{Buzdin1991}  Hence it is plausible that simply changing the electronic transport regime inside the junction from ballistic to diffusive could account for the shift observed by Baek \textit{et al.}  Alternatively, it is known that adding potential barriers at interfaces may also shift the position of the 0-$\pi$ transition,\cite{Faure2006,Heim2015} so this issue is by no means resolved.

In any case, there is no published theoretical formula that can account for all the complexities of the spin-valve junctions used in the experiments.  In particular, the mismatch of Fermi surface properties at the various interfaces inside the junctions is an issue that demands a detailed microscopic calculation.  In order to make progress, we utilize the formulas derived by Crouzy, Tollis, and Ivanov,\cite{Crouzy2007} which predict the critical currents in both the P and AP magnetic states for junctions containing two F layers, in the diffusive limit:
	\begin{equation}
	\label{eqn:CrouzyP}
	I_c^P = I_0 Re[\frac{(1+i)}{(\textrm{sinh}[(1+i)(d_{F1}/\xi_{F1}+d_{F2}/\xi_{F2})])}],
	\end{equation}
	\begin{equation}
	\label{eqn:CrouzyAP}
    I_c^{AP} = I_0 Re[\frac{2}{\textrm{sin}[(1+i)d_{F1}/\xi_{F1}+(1-i)d_{F2}/\xi_{F2}]+\textrm{sinh}[(1-i)d_{F1}/\xi_{F1}+(1+i)d_{F2}/\xi_{F2}]}]
	\end{equation}
Using those formulas, the appearance of a $\pi$ state is manifested by a negative critical current.  In the left side of Figure 1 we plot a theoretical ``phase diagram" of 0 and $\pi$ states for both the P and AP magnetic states.  This figure is very similar to Figure 2 in the paper by Crouzy \textit{et al.},\cite{Crouzy2007} except that the value of $\xi_F$ may be different for the two magnetic materials.  In Figure 1, the dark yellow lines with negative slope indicate transitions between 0 and $\pi$ states for the P magnetic state, with the state in the lower left-hand corner being the 0 state.  The blue lines with positive slope indicate transitions between 0 and $\pi$ states for the AP magnetic state.  Each box in Figure 1 is labeled with the phase states of a junction in the P state and AP state.  Useful devices are located in boxes where the two phase states are different, i.e. either ``0-$\pi$'' or ``$\pi$-0''.

   \begin{figure}[ht]
   \begin{center}
   \begin{tabular}{c} 
   \includegraphics[height=6cm]{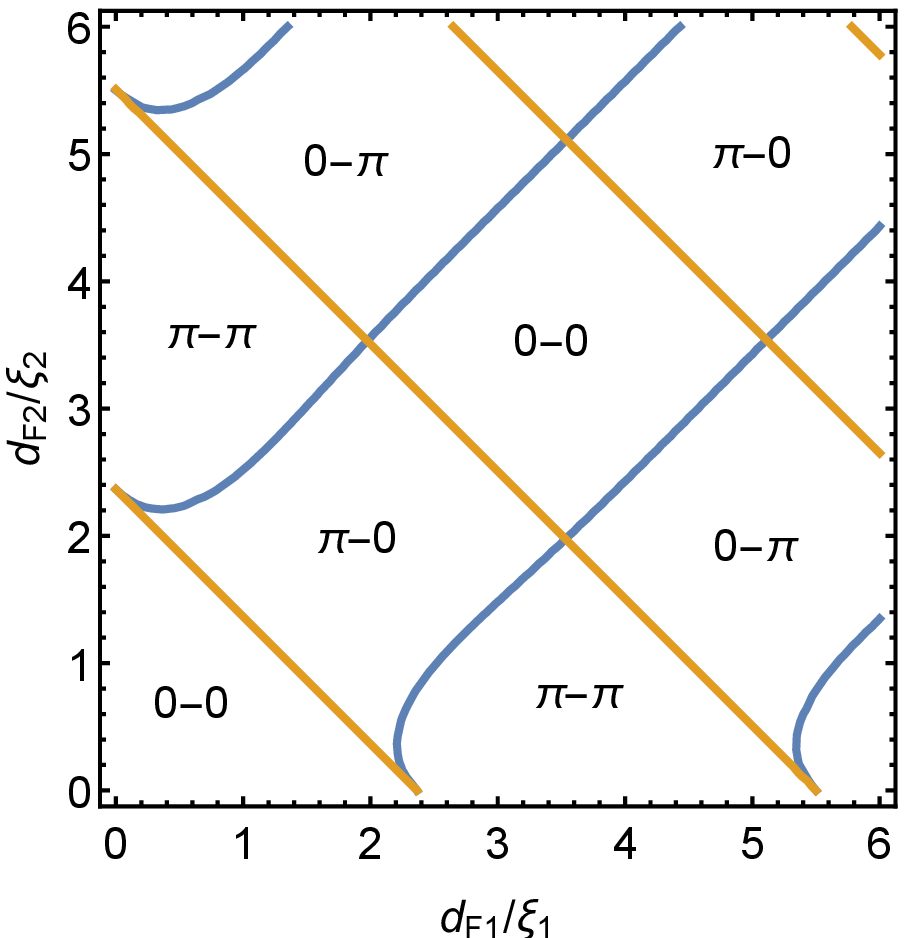}
   \includegraphics[height=6cm]{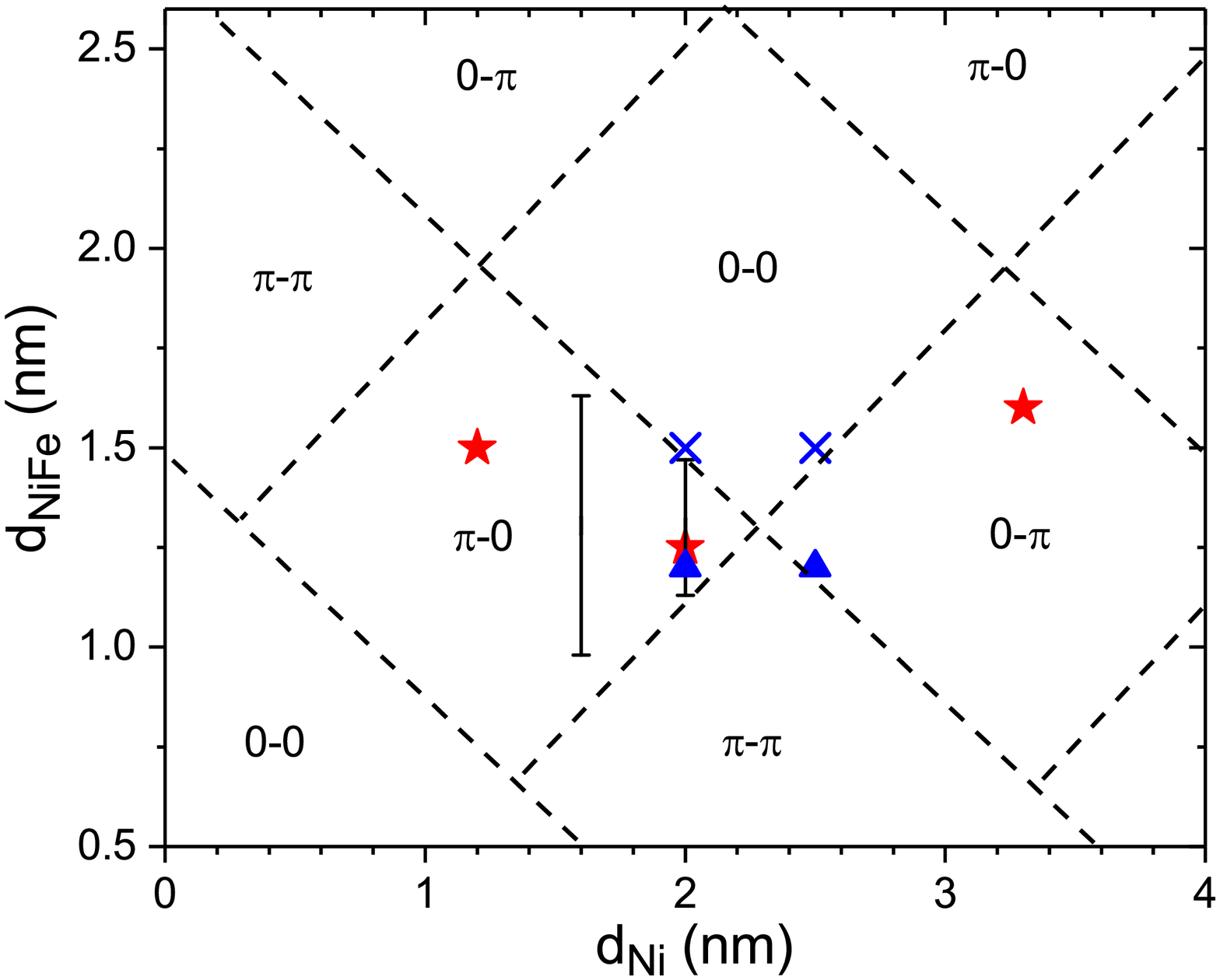}
   \end{tabular}
   \end{center}
   \caption[example]
   { \label{fig:CrouzyDiagram}
Tentative phase diagram of spin-valve Josephson junctions containing a Ni fixed layer and NiFe free layer.  Left: theoretical diagram based on calculations in Ref.~\citenum{Crouzy2007}, modified in Eqns. (1) and (2) to account for two different ferromagnetic materials with different values of $\xi_F$.  The dark yellow lines with negative slope separate regions of 0 and $\pi$ phase states in the parallel (P) magnetic state.  The blue lines with mostly positive slope do the same for the antiparallel (AP) magnetic state.  Right: experimental phase diagram based on the data in the Table.  Red stars represent published data from devices that have been demonstrated to exhibit controllable switching between the 0 and $\pi$ phase states.  The vertical bars represent measurements on single junctions, as described in the text and Table.  Blue triangles and crosses represent unpublished data from devices that did or did not exhibit phase control, respectively.  The dashed lines represent our best guess at the phase boundaries based on the arguments in the text and a comparison with the figure on the left.  (We did not attempt to draw the curved ends of the dashed lines where they approach the axes.)  In both figures each box is labeled with the phase states of a Josephson junction in the P and AP magnetic states, respectively.
}
   \end{figure}

The lines in the left side of Figure 1 have simple interpretations.  In the P state, the junction behaves as though it contains only a single F layer with total reduced thickess $d_{F1}/\xi_{F1}+d_{F2}/\xi_{F2}$.  In the diffusive limit, such a junction has 0-$\pi$ transitions whenever the reduced thickness is equal to $(n+3/4)\pi$, with $n$ a non-negative integer.  Thus the yellow lines can be reproduced by the simple equation:

 	\begin{equation}
	\label{eqn:CrouzyPbounaries}
	d_{F1}/\xi_{F1}+d_{F2}/\xi_{F2}=(n+3/4)\pi
	\end{equation}

In the AP state, the situation is more complicated.  Far from the axes, the junction behaves as though the phase accumulation of an electron pair is equal to the difference between the phase accumulation in each F layer.  The extra phase shift of $\pi/4$ from each layer cancels, hence the straight parts of the blue lines can be reproduced by:

 	\begin{equation}
	\label{eqn:CrouzyAPbounaries}
	d_{F1}/\xi_{F1}-d_{F2}/\xi_{F2}=(n+1/2)\pi
	\end{equation}

That simple formula fails, however, when the thickness of one of the layers approaches zero.  In that limit we know that the 0-$\pi$ phase boundary depends only on the thickness of the other layer, and it occurs when the reduced layer thickness is $3\pi/4$ rather than $\pi/2$.  Hence the blue lines curve near the axes to accommodate the appearance of the extra $\pi/4$ phase shift.

Now we consider all the data published to date that shed light on the phase behavior of spin-valve Josephson junctions containing a Ni fixed layer and a NiFe free layer.  Three entries in the Table below represent phase-sensitive measurements from references \citenum{Gingrich2016,Dayton2018,Madden2018}.  Each of those works presents data from junctions with a single specific value of the Ni thickness and NiFe thickness in the junctions.  In addition, Ref. \citenum{Niedzielski2018} reports critical current amplitude measurements on junctions with two different Ni thicknesses, while varying the NiFe thickness over a substantial range.  Those measurements were not phase-sensitive; nevertheless, they indicated at what NiFe thicknesses the critical current passed through zero in both the P and AP states, which we interpret as phase boundaries between 0 and $\pi$ states.  It then requires at least one phase-sensitive measurement to nail down which region of the phase diagram corresponds to functional devices where the P and AP states correspond to different phase states.  The work by Madden \textit{et al.}\cite{Madden2018} confirmed the interpretation in Ref.~\citenum{Niedzielski2018} that junctions with NiFe thicknesses in the range given in the table do indeed correspond to functional devices.  The last two lines of the table contain unpublished results from NGC.  With a NiFe thickness of 1.2 nm, NGC workers observed phase control with Ni thicknesses of both 2.0 and 2.5 nm.  When the NiFe thickness was increased to 1.5 nm, however, they did not observe any phase change between the P and AP magnetic states.  From the amplitudes of the critical current in the P and AP states, the NGC group guessed that all four types of samples lay close to a phase boundary.

\begin{table}[ht]
\caption{Known thicknesses that do or do not produce controllable 0-$\pi$ junctions.}
\label{tab:Thicknesses}
\begin{center}
\begin{tabular}{|l|l|l|}
\hline
\rule[-1ex]{0pt}{3.5ex}  $d_{Ni}$ (nm) & $d_{NiFe}$ (nm) & Comments  \\
\hline
\rule[-1ex]{0pt}{3.5ex}  1.2 & 1.5 cm & Demonstrated phase control, Ref. \citenum{Gingrich2016}   \\
\hline
\rule[-1ex]{0pt}{3.5ex}  1.6 & 1.0 - 1.6 & Amplitude measurement only, Ref. \citenum{Niedzielski2018}  \\
\hline
\rule[-1ex]{0pt}{3.5ex}  2.0 & 1.1 - 1.5 & Amplitude measurement only, Ref. \citenum{Niedzielski2018}  \\
\hline
\rule[-1ex]{0pt}{3.5ex}  2.0 & 1.25 cm & Demonstrated phase control, Ref. \citenum{Madden2018}  \\
\hline
\rule[-1ex]{0pt}{3.5ex}  3.3 & 1.6 cm & Demonstrated phase control, Ref. \citenum{Dayton2018}  \\
\hline
\rule[-1ex]{0pt}{3.5ex}  2.0, 2.5 & 1.2 cm & Demonstrated phase control, NGC unpublished data  \\
\hline
\rule[-1ex]{0pt}{3.5ex}  2.0, 2.5 & 1.5 cm & No phase control, NGC unpublished data  \\
\hline
\end{tabular}
\end{center}
\end{table}

The data in the table are summarized in the right side of Figure 1.  Each of the three red stars corresponds to one of the published phase-sensitive measurements.  The vertical lines correspond to the amplitude measurements.  The blue symbols represent the unpublished data from NGC -- triangles for successes and crosses for failures.  We now attempt to place lines on the right side of Figure 1 following roughly the scheme indicated in the theoretical picture on the left side.  Three types of information are used to guide us.  First, each red star or blue triangle must be placed inside a region corresponding to functional devices, while the blue crosses should lie in a region where the P and AP states have the same phase.  Second, the ends of the vertical lines should lie close to phase boundaries.  Third, we have additional information regarding the behavior of the lines near the axes, from junctions containing only a single F layer.  In junctions containing only Ni, Baek \textit{et al.} showed that the junctions exhibit 0-$\pi$ transitions at Ni thicknesses of approximately 0.8 and 3.3 nm, which correspond well to the theoretical formula for the ballistic limit.\cite{Baek2018}  When diffusing layers are added to those junctions, however, the 0-$\pi$ transition thicknesses change to about 1.4 and 3.7 nm.  We therefore expect that the phase-boundary lines should intersect the horizontal axis representing Ni thickness at approximately 1.4 and 3.7 nm.  In junctions containing only NiFe, Glick \textit{et al.}\cite{Glick2017} showed that junctions exhibit 0-$\pi$ transitions at NiFe thicknesses of approximately 1.7 and 3.5 nm, so we expect the phase-boundary lines to intersect the vertical axis at approximately those values.  If we now sit down with pencil and paper and try to draw straight phase boundary lines that meet all of these requirements, we quickly realize that it is impossible.  So we have relaxed the criteria somewhat and come up with the lines shown in the figure.  These lines satisfy the first criterion (stars and triangles appear only in viable regions, crosses do not), they approximately satisfy the second criterion (end of vertical lines lie close to phase boundaries), but they violate somewhat the third criterion (lines meet the axes at the known single-junction phase boundary points).  The astute reader will notice that the vertical axis does not start at zero, but rather at 0.5 nm.  That was done to aid in making the experimental phase diagram on the right side of Figure 1 look similar to the theoretical diagram on the left side, and might be justified by claiming that NiFe has a magnetically ``dead" layer that does not contribute to the electron pair phase shift in Josephson junctions.  We have no proof of that statement; but trying to draw approximate phase boundary lines in Figure 1 while starting the vertical axis at zero proved impossible.  Needless to say, if we allowed the phase boundary lines to curve, then it would become much easier to satisfy all the simultaneous constraints.  We have refrained from doing that at this time, given the paucity of available data to fully constrain the curves.

We present this phase diagram not as a definitive statement, but more as a guide for future research.  For example, we have determined that the magnetic properties of Ni tend to improve with increasing thickness.  Starting with the Ni thickness of 2.0 nm studied in Ref. \citenum{Madden2018}, it would be tempting to increase the thickness slightly, say to 2.4 nm, to see if the magnetic behavior improves.  According to the phase diagram, however, such a move would be quite disappointing, as there would be a vanishingly small range of NiFe thicknesses that produce viable devices at that Ni thickness.  A similar argument could be made against using a NiFe thickness of 2.0 nm, although we note that there is much more uncertainty about the phase boundary locations along that axis.

\section{Conclusions}

The development of a large-scale superconducting memory is still in its infancy.  This paper has only discussed one particular form of such memory, the Josephson magnetic random access memory based on phase-controllable ferromagnetic Josephson junctions.\cite{Dayton2018}  In addition to giving a snapshot of where that technology stands at the moment, we have provided a speculative phase diagram of useful devices for a particular choice of magnetic materials, namely Ni and NiFe.  The diagram is based on a theory that is known to be incomplete; our hope is that such a diagram will be useful both to guide experimentalists and to spur the development of more realistic theories that could explain all of the experimental data in a rigorous way.

\acknowledgments 

This paper is based on work carried out with a large number of collaborators, including many former graduate and undergraduate students.  The students whose work influenced this paper most directly include E.C. Gingrich, J.A. Glick, and B.M. Niedzielski.  We thank T.F. Ambrose, I.M. Dayton, and M.G. Loving for their work on the NGC devices.  We also have benefitted from numerous discussions with A.Y. Herr, D. Miller, N. Newman, N.D. Rizzo, and W.P. Pratt, Jr.  This research is based upon work supported by the ODNI, IARPA, via ARO contract number W911NF-14-C-0115. The views and conclusions contained herein are those of the authors and should not be interpreted as necessarily representing the official policies or endorsements, either expressed or implied, of the ODNI, IARPA, or the U.S. Government.

\bibliography{Birge_SPIE_bib_v5} 

\begin{thebibliography}{10}

\bibitem{Ryazanov2001}
Ryazanov, V.~V., Oboznov, V.~A., Rusanov, A.~Y., Veretennikov, A.~V., Golubov,
  A.~A., and Aarts, J., ``Coupling of two superconductors through a
  ferromagnet: Evidence for a $\pi$ junction,'' {\em Phys. Rev. Lett.}~{\bf
  86},  2427--2430 (Mar 2001).

\bibitem{Kontos2002}
Kontos, T., Aprili, M., Lesueur, J., Gen{\^e}t, F., Stephanidis, B., and
  Boursier, R., ``{Josephson Junction through a Thin Ferromagnetic Layer:
  Negative Coupling},'' {\em Phys. Rev. Lett.}~{\bf 89},  137007 (Sep 2002).

\bibitem{Buzdin1982}
Buzdin, A.~I., Bulaevskii, L.~N., and Panyukov, S.~V., ``{Critical-current
  oscillations as a function of the exchange field and thickness of the
  ferromagnetic metal (F) in an S-F-S Josephson junction},'' {\em Pis'ma Zh.
  Eksp. Teor. Fiz.}~{\bf 35}(4),  147--148 (1982).
\newblock [\textit{J. Exp. Theor. Phys. Lett.}, \textbf{35}, 4, 20 (1982)].

\bibitem{Buzdin1991}
Buzdin, A.~I. and Kupriyanov, M.~Y., ``Josephson junction with a ferromagnetic
  layer,'' {\em Pis'ma Zh. Eksp. Teor. Fiz.}~{\bf 53}(6),  308--312 (1991).
\newblock [\textit{J. Exp. Theor. Phys. Lett.}, \textbf{53}, 6, 321 (1991)].

\bibitem{Faris1980}
Faris, S.~M., Henkels, W.~H., Valsamakis, E.~A., and Zappe, H.~H., ``Basic
  design of a {Josephson} technology cache memory,'' {\em IBM J. Res.
  Dev.}~{\bf 24},  143--154 (March 1980).

\bibitem{Miyahara1984}
Miyahara, K., Yamauchi, Y., Yamamoto, M., and Ishida, A., ``An improved {NDRO
  Josephson} quantized loop memory cell with buffering configuration,'' {\em
  IEEE Trans. Electron Devices}~{\bf 31},  888--894 (July 1984).

\bibitem{LikharevSemenov1991}
Likharev, K. and Semenov, V., ``{RSFQ Logic/Memory Family: A New
  Josephson-Junction Technology for Sub-Terahertz-Clock-Frequency Digital
  Systems},'' {\em IEEE Trans. Appl. Supercond.}~{\bf 50}(1) (1991).

\bibitem{Nagasawa1995}
Nagasawa, S., Hashimoto, Y., Numata, H., and Tahara, S., ``{A 380 ps, 9.5 mW
  Josephson 4-Kbit RAM operated at a high bit yield},'' {\em IEEE Trans. Appl.
  Supercond.}~{\bf 5},  2447--2452 (June 1995).

\bibitem{Kirichenko2011}
Kirichenko, D.~E., Sarwana, S., and Kirichenko, A.~F., ``Zero static power
  dissipation biasing of {RSFQ} circuits,'' {\em IEEE Trans. Appl.
  Supercond.}~{\bf 21},  776--779 (June 2011).

\bibitem{Volkmann2013}
Volkmann, M.~H., Sahu, A., Fourie, C.~J., and Mukhanov, O.~A., ``Implementation
  of energy efficient single flux quantum digital circuits with {sub-aJ/bit}
  operation,'' {\em Supercond. Sci. Technol.}~{\bf 26}(1),  015002 (2013).

\bibitem{Herr2011}
Herr, Q.~P., Herr, A.~Y., Oberg, O.~T., and Ioannidis, A.~G., ``Ultra-low-power
  superconductor logic,'' {\em J. Appl. Phys.}~{\bf 109}(10),  103903 (2011).

\bibitem{Tanaka2012}
Tanaka, M., Ito, M., Kitayama, A., Kouketsu, T., and Fujimaki, A., ``{18-GHz,
  4.0-aJ/bit} operation of ultra-low-energy rapid single-flux-quantum shift
  registers,'' {\em Japanese J. Appl. Phys.}~{\bf 51}(5R),  053102 (2012).

\bibitem{Takeuchi2013}
Takeuchi, N., Ozawa, D., Yamanashi, Y., and Yoshikawa, N., ``An adiabatic
  quantum flux parametron as an ultra-low-power logic device,'' {\em Supercond.
  Sci. Technol.}~{\bf 26}(3),  035010 (2013).

\bibitem{Holmes2013}
Holmes, D.~S., Ripple, A.~L., and Manheimer, M.~A., ``{Energy-efficient
  superconducting computing---power budgets and requirements},'' {\em IEEE
  Trans. Appl. Supercond.}~{\bf 23},  1701610 (2013).

\bibitem{Soloviev2017_review}
Soloviev, I.~I., Klenov, N.~V., Bakurskiy, S.~V., Kupriyanov, M.~Y., Gudkov,
  A.~L., and Sidorenko, A.~S., ``Beyond {M}oore’s technologies: operation
  principles of a superconductor alternative,'' {\em Beilstein J.
  Nanotechnol.}~{\bf 8},  2689 (2017).

\bibitem{Mannhart2006}
Held, R., Xu, J., Schmehl, A., Schneider, C.~W., Mannhart, J., and Beasley,
  M.~R., ``Superconducting memory based on ferromagnetism,'' {\em Appl. Phys.
  Lett.}~{\bf 89}(16),  163509 (2006).

\bibitem{Johnson1999}
Clinton, T.~W. and Johnson, M., ``Nonvolatile switchable {Josephson}
  junctions,'' {\em J. Appl. Phys.}~{\bf 85}(3),  1637--1643 (1999).

\bibitem{Beasley1997}
Oh, S., Youm, D., and Beasley, M.~R., ``A superconductive magnetoresistive
  memory element using controlled exchange interaction,'' {\em Appl. Phys.
  Lett.}~{\bf 71}(16),  2376--2378 (1997).

\bibitem{Bell2004}
Bell, C., Burnell, G., Leung, C.~W., Tarte, E.~J., Kang, D.-J., and Blamire,
  M.~G., ``Controllable {Josephson} current through a pseudospin-valve
  structure,'' {\em Appl. Phys. Lett.}~{\bf 84}(7),  1153--1155 (2004).

\bibitem{Leksin2010}
Leksin, P.~V., Garifyanov, N.~N., Garifullin, I.~A., Schumann, J., Vinzelberg,
  H., Kataev, V., Klingeler, R., Schmidt, O.~G., and Buchner, B., ``Full spin
  switch effect for the superconducting current in a superconductor/ferromagnet
  thin film heterostructure,'' {\em Appl. Phys. Lett.}~{\bf 97}(10),  102505
  (2010).

\bibitem{Lyuksyutov2005}
Lyuksyutov, I. and Pokrovsky, V., ``Ferromagnet-superconductor hybrids,'' {\em
  Advances in Physics}~{\bf 54}(1),  67--136 (2005).

\bibitem{Buzdin2005}
Buzdin, A.~I., ``{Proximity effects in superconductor-ferromagnet
  heterostructures},'' {\em Rev. Mod. Phys.}~{\bf 77},  935--976 (Sep 2005).

\bibitem{Bergeret2005}
Bergeret, F.~S., Volkov, A.~F., and Efetov, K.~B., ``Odd triplet
  superconductivity and related phenomena in superconductor-ferromagnet
  structures,'' {\em Rev. Mod. Phys.}~{\bf 77},  1321--1373 (Nov 2005).

\bibitem{Demler1997}
Demler, E.~A., Arnold, G.~B., and Beasley, M.~R., ``Superconducting proximity
  effects in magnetic metals,'' {\em Phys. Rev. B}~{\bf 55},  15174--15182 (Jun
  1997).

\bibitem{Frolov2004}
Frolov, S.~M., Van~Harlingen, D.~J., Oboznov, V.~A., Bolginov, V.~V., and
  Ryazanov, V.~V., ``Measurement of the current-phase relation of
  superconductor/ferromagnet/superconductor $\ensuremath{\pi}$ {Josephson}
  junctions,'' {\em Phys. Rev. B}~{\bf 70},  144505 (Oct 2004).

\bibitem{Bergeret2001PRL}
Bergeret, F.~S., Volkov, A.~F., and Efetov, K.~B., ``Enhancement of the
  {Josephson} current by an exchange field in superconductor-ferromagnet
  structures,'' {\em Phys. Rev. Lett.}~{\bf 86},  3140--3143 (Apr 2001).

\bibitem{Krivoruchko2001}
Krivoruchko, V.~N. and Koshina, E.~A., ``{From inversion to enhancement of the
  dc Josephson current in $S/F\ensuremath{-}I\ensuremath{-}F/S$ tunnel
  structures},'' {\em Phys. Rev. B}~{\bf 64},  172511 (Oct 2001).

\bibitem{Golubov2002}
Golubov, A.~A., Kupriyanov, M.~Y., and Fominov, Y.~V., ``Critical current in
  {SFIFS} junctions,'' {\em JETP Lett.}~{\bf 75},  190--194 (Feb 2002).

\bibitem{herr_patent2012}
Herr, A.~Y. and Herr, Q.~P., ``{Josephson magnetic random access memory system
  and method},'' (Sept.~18 2012).
\newblock $\mathrm{US}$ Patent 8,270,209.

\bibitem{herr_phasepatent2015}
Herr, A., Herr, Q., and Naaman, O., ``{Phase hysteretic magnetic Josephson
  junction memory cell},'' (Apr.~2 2015).
\newblock $\mathrm{US}$ Patent 9,208,861.

\bibitem{Dayton2018}
Dayton, I.~M., Sage, T., Gingrich, E.~C., Loving, M.~G., Ambrose, T.~F., Siwak,
  N.~P., Keebaugh, S., Kirby, C., Miller, D.~L., Herr, A.~Y., Herr, Q.~P., and
  Naaman, O., ``Experimental demonstration of a {Josephson} magnetic memory
  cell with a programmable $\pi$-junction,'' {\em IEEE Magn. Lett.}~{\bf 9},
  1--5 (2018).

\bibitem{LinderRobinson2015}
Linder, J. and Robinson, J. W.~A., ``Superconducting spintronics,'' {\em Nat.
  Phys.}~{\bf 11},  307--315 (Apr 2015).
\newblock Review.

\bibitem{Eschrig2015}
Eschrig, M., ``Spin-polarized supercurrents for spintronics: a review of
  current progress,'' {\em Rep. Prog. Phys.}~{\bf 78}(10),  104501 (2015).

\bibitem{Glick2018}
Glick, J.~A., Aguilar, V., Gougam, A.~B., Niedzielski, B.~M., Gingrich, E.~C.,
  Loloee, R., Pratt, W.~P., and Birge, N.~O., ``Phase control in a spin-triplet
  {SQUID},'' {\em Science Advances}~{\bf 4}(7) (2018).

\bibitem{Gingrich2016}
Gingrich, E.~C., Niedzielski, B.~M., Glick, J.~A., Wang, Y., Miller, D.~L.,
  Loloee, R., {Pratt Jr}, W.~P., and Birge, N.~O., ``{Controllable 0-$\pi$
  {Josephson} junctions containing a ferromagnetic spin valve},'' {\em Nat.
  Phys.}~{\bf 12},  564--567 (Jun 2016).

\bibitem{Wang2012}
Wang, Y., {Pratt Jr}, W.~P., and Birge, N.~O., ``{Area-dependence of
  spin-triplet supercurrent in ferromagnetic Josephson junctions},'' {\em Phys.
  Rev. B}~{\bf 85},  214522 (2012).

\bibitem{Thomas1998}
Thomas, C.~D., Ulmer, M.~P., and Ketterson, J.~B., ``Superconducting tunnel
  junction base electrode planarization,'' {\em J. App. Phys.}~{\bf 84}(1),
  364--367 (1998).

\bibitem{Singh2018}
Singh, R.~K., Rizzo, N.~D., Bertram, M., Zheng, K., and Newman, N.,
  ``Improvement in the magnetic properties of {Ni-Fe} thin films on thick {Nb}
  electrodes using oxidation and low-energy {Ar} ion milling,'' {\em IEEE Magn.
  Lett.}~{\bf 9},  1--4 (2018).

\bibitem{Loving2018}
Loving, M.~G., Ambrose, T.~F., Ermer, H., Miller, D., and Naaman, O.,
  ``Interplay between interface structure and magnetism in {NiFe/Cu/Ni}-based
  pseudo-spin valves,'' {\em AIP Advances}~{\bf 8}(5),  056309 (2018).

\bibitem{Ryazanov2012}
Ryazanov, V.~V., Bol'ginov, V.~V., Sobanin, D.~S., Vernik, I.~V., Tolpygo,
  S.~K., Kadin, A.~M., and Mukhanov, O.~A., ``Magnetic josephson junction
  technology for digital and memory applications,'' {\em Physics Procedia}~{\bf
  36},  35 -- 41 (2012).
\newblock SUPERCONDUCTIVITY CENTENNIAL Conference 2011.

\bibitem{Glick2017PdFe}
Glick, J.~A., Loloee, R., Pratt, W.~P., and Birge, N.~O., ``Critical current
  oscillations of {Josephson} junctions containing {PdFe} nanomagnets,'' {\em
  IEEE Trans. Appl. Supercond.}~{\bf 27},  1--5 (June 2017).

\bibitem{Blum2002}
Blum, Y., Tsukernik, A., Karpovski, M., and Palevski, A., ``Oscillations of the
  superconducting critical current in {Nb-Cu-Ni-Cu-Nb} junctions,'' {\em Phys.
  Rev. Lett.}~{\bf 89},  187004 (Oct 2002).

\bibitem{Shelukhin2006}
Shelukhin, V., Tsukernik, A., Karpovski, M., Blum, Y., Efetov, K.~B., Volkov,
  A.~F., Champel, T., Eschrig, M., L\"ofwander, T., Sch\"on, G., and Palevski,
  A., ``Observation of periodic $\ensuremath{\pi}$-phase shifts in
  ferromagnet-superconductor multilayers,'' {\em Phys. Rev. B}~{\bf 73},
  174506 (May 2006).

\bibitem{Robinson2006}
Robinson, J. W.~A., Piano, S., Burnell, G., Bell, C., and Blamire, M.~G.,
  ``{Critical current oscillations in strong ferromagnetic $\pi$ junctions},''
  {\em Phys. Rev. Lett.}~{\bf 97},  177003 (2006).

\bibitem{Glick2017}
Glick, J.~A., Khasawneh, M.~A., Niedzielski, B.~M., Loloee, R., Pratt, W.~P.,
  Birge, N.~O., Gingrich, E.~C., Kotula, P.~G., and Missert, N., ``Critical
  current oscillations of elliptical {Josephson} junctions with single-domain
  ferromagnetic layers,'' {\em J. Appl. Phys.}~{\bf 122}(13),  133906 (2017).

\bibitem{Baek2014}
Baek, B., Rippard, W.~H., Benz, S.~P., Russek, S.~E., and Dresselhaus, P.~D.,
  ``{Hybrid superconducting-magnetic memory device using competing order
  parameters},'' {\em Nature Commun.}~{\bf 5},  3888 (2014).

\bibitem{Niedzielski2015}
Niedzielski, B.~M., Gingrich, E.~C., Loloee, R., Pratt, W.~P., and Birge,
  N.~O., ``{S/F/S} {Josephson} junctions with single-domain ferromagnets for
  memory applications,'' {\em Supercond. Sci. Technol.}~{\bf 28}(8),  085012
  (2015).

\bibitem{Bethanythesis}
Niedzielski, B.~M., {\em Phase Sensitive Measurements of Ferromagnetic
  Josephson Junctions for Cryogenic Memory Applications}, PhD thesis, Michigan
  State University (2017).

\bibitem{Oboznov2006}
Oboznov, V.~A., Bol'ginov, V.~V., Feofanov, A.~K., Ryazanov, V.~V., and Buzdin,
  A.~I., ``{Thickness Dependence of the Josephson Ground States of
  Superconductor-Ferromagnet-Superconductor Junctions},'' {\em Phys. Rev.
  Lett.}~{\bf 96},  197003 (May 2006).

\bibitem{Niedzielski2014}
Niedzielski, B.~M., Diesch, S.~G., Gingrich, E.~C., Wang, Y., Loloee, R.,
  Pratt, W.~P., and Birge, N.~O., ``Use of {Pd-Fe} and {Ni-Fe-Nb} as soft
  magnetic layers in ferromagnetic {Josephson} junctions for nonvolatile
  cryogenic memory,'' {\em IEEE Trans. Appl. Supercond.}~{\bf 24},  1--7 (Aug
  2014).

\bibitem{Mukhanov2012}
Vernik, I.~V., Bol'ginov, V.~V., Bakurskiy, S.~V., Golubov, A.~A., Kupriyanov,
  M.~Y., Ryazanov, V.~V., and Mukhanov, O.~A., ``Magnetic josephson junctions
  with superconducting interlayer for cryogenic memory,'' {\em IEEE Trans.
  Appl. Supercond.}~{\bf 23},  1701208--1701208 (June 2013).

\bibitem{Baek2017}
Baek, B., Schneider, M.~L., Pufall, M.~R., and Rippard, W.~H., ``Phase offsets
  in the critical-current oscillations of {Josephson} junctions based on {Ni
  and
  Ni-${\mathbf{(}{\text{Ni}}_{81}{\text{Fe}}_{19}\mathbf{)}}_{x}{\mathrm{Nb}}_{y}$}
  barriers,'' {\em Phys. Rev. Applied}~{\bf 7},  064013 (Jun 2017).

\bibitem{Niedzielski2018}
Niedzielski, B.~M., Bertus, T.~J., Glick, J.~A., Loloee, R., Pratt, W.~P., and
  Birge, N.~O., ``Spin-valve {Josephson} junctions for cryogenic memory,'' {\em
  Phys. Rev. B}~{\bf 97},  024517 (Jan 2018).

\bibitem{Faure2006}
Faur{\'e}, M., Buzdin, A.~I., Golubov, A.~A., and Kupriyanov, M.~Y.,
  ``{Properties of superconductor/ferromagnet structures with spin-dependent
  scattering},'' {\em Phys. Rev. B}~{\bf 73},  064505 (Feb 2006).

\bibitem{Heim2015}
Heim, D.~M., Pugach, N.~G., Kupriyanov, M.~Y., Goldobin, E., Koelle, D.,
  Kleiner, R., Ruppelt, N., Weides, M., and Kohlstedt, H., ``{The effect of
  normal and insulating layers on 0-$\pi$ transitions in Josephson junctions
  with a ferromagnetic barrier},'' {\em New J. Phys.}~{\bf 17}(11),  113022
  (2015).

\bibitem{Crouzy2007}
Crouzy, B., Tollis, S., and Ivanov, D.~A., ``Josephson current in a
  superconductor-ferromagnet junction with two noncollinear magnetic domains,''
  {\em Phys. Rev. B}~{\bf 75},  054503 (Feb 2007).

\bibitem{Madden2018}
Madden, A., Willard, J., Loloee, R., and Birge, N.~O., ``Phase controllable
  {Josephson} junctions for cryogenic memory,'' {\em arXiv:1808.02914}  (2018).

\bibitem{Baek2018}
Baek, B., Schneider, M.~L., Pufall, M.~R., and Rippard, W.~H., ``Anomalous
  supercurrent modulation in {Josephson} junctions with {Ni}-based barriers,''
  {\em IEEE Trans. Appl. Supercond.}~{\bf 28},  1--5 (Oct 2018).

\end{thebibliography}
\bibliographystyle{spiebib} 

\end{document}